\documentclass[5p]{elsarticle}

\usepackage{graphics}
\usepackage{eufrak}
\usepackage{amsmath,amssymb}
\usepackage{amsmath,color}
\usepackage{graphicx, hyperref}
\usepackage{psfrag}
\usepackage{textcomp}
\usepackage[usenames,dvipsnames]{xcolor}
\usepackage{ulem}

\biboptions{sort&compress}

\newcommand{\be}{\begin{equation}}
\newcommand{\ee}[1]{\label{ #1} \end{equation}}

\newcommand{\pd}[2]{\frac{\partial #1}{\partial #2} }

\begin{document}
\title{\bf Volume dependent extension of Kerr-Newman black hole thermodynamics}

\author[wigner]{Tam\'as S.~Bir\'o}
\ead{biro.tamas@wigner.mta.hu}
\author[wigner]{Viktor G.~Czinner}
\ead{czinner.viktor@wigner.mta.hu}
\author[nihon]{Hideo Iguchi\corref{mycorrespondingauthor}}
\cortext[mycorrespondingauthor]{Corresponding author}
\ead{iguchi.h@phys.ge.cst.nihon-u.ac.jp}
\author[wigner,bme]{P\'eter V\'an}
\ead{van.peter@wigner.mta.hu}

\address[wigner]{Wigner Research Centre for Physics, H-1525 Budapest, P.O.Box 49, Hungary}
\address[nihon]{Laboratory of Physics, College of Science and Technology, Nihon University, 
274-8501 Narashinodai, Funabashi, Chiba, Japan}
\address[bme]{Department of Energy Engineering, BME Faculty of Mech. Eng.,
Bertalan Lajos u.~4-6, 1111 Budapest, Hungary}

\begin{abstract}
{%
	We show that the Hawking--Bekenstein entropy formula is modified by a factor of $8/3$ 
	if one also considers a work term in the 1st law of thermodynamics by a pressure stemming 
	from the Hawking radiation. We give an intuitive definition for the corresponding thermodynamical volume
	by the implicit definition $\epsilon=Mc^2/V$, which is the average energy density of the Hawking radiation. This volume scales  as $V \sim M^5$, agreeing with other suggestions. 
	As a result the corresponding Smarr relation describes an extensive entropy
	and 
	a stable effective equation of state $S(E,V)\sim E^{3/4}V^{1/4}$.
	These results pertain for charged and rotating Kerr-Newman black holes.}
\end{abstract}

\begin{keyword}
Black holes \sep  Volume \sep Entropy \sep Thermodynamics \sep Heat capacity \sep Thermal stability

\end{keyword}

\date{\today}

\maketitle
\section{Introduction}
\label{sec:Introduction}

Most paradox features of black hole thermodynamics are due to the seemingly convex nature of the entropy function. 
The Bekenstein-Hawking entropy  \cite{Bekenstein:1972tm,Bekenstein:1973ur,Hawking:1974sw} was originally derived to 
be proportional to the area of the event horizon,  and since the horizon radius of a Schwarzschild black hole 
is proportional to its mass, 
the entropy is apparently a convex function with respect to the mass-energy parameter ($E \approx Mc^2$). 
In this case the heat capacity, $C$, related to the second derivative of the $S(E)$ function via $S''(E)=-1/CT^2$, 
is negative.
Furthermore, this result of proportionality to the area violates the extensivity of the entropy  function. 
By considering a generalized nonadditive entropy definition, a possible solution to the problem was investigated in \cite{Tsallis:2012js}.
Later on, it was also argued that the convexity of the black hole entropy can be removed by a R\'enyi entropy based 
consideration \cite{Biro:2013cra,Czinner:2015eyk,Czinner:2017tjq}.

As an independent direction, the physical volume of a black hole has long been discussed by various authors 
\cite{Parikh:2005qs,Grumiller:2005zk,Ballik:2010rx,Ballik:2013uia,DiNunno:2008id,Finch:2012fq,Cvetic:2010jb,Gibbons:2012ac}.
{
In \cite{Kastor:2009wy} {for example}, the negative cosmological constant is treated as pressure of AdS black holes, and 
an effective volume is introduced as its conjugate variable. Thereby the mass of an AdS black hole can be considered as the enthalpy of the spacetime.}
More recently, Christodoulou and Rovelli proposed a geometric invariant definition, where the volume of the 
Schwarz\-schild back hole is defined as the largest, spherically symmetric, spacelike hypersurface bounded by the 
event horizon \cite{Christodoulou:2014yia}.
The corresponding CR-volume can become surprisingly large, it scales as $V \sim M^5$ when the thermal property of 
the Hawking radiation is taken into account. Soon after this definition, the CR approach
has been extended to rotating black holes \cite{Bengtsson:2015zda,Wang:2019ake}, 
and it also motivated several authors to calculate the entropy corresponding to a 
black hole with volume \cite{Zhang:2015gda,Majhi:2017tab,Zhang:2017aqf,Han:2018jnf,Wang:2018dvo}, and investigate the role of the volume in their thermodynamic behavior \cite{Christodoulou:2016tuu,Ong:2015tua,Ong:2015dja,Rovelli:2017mzl}.  These investigations, however, mainly consider the contribution of a hypothetic scalar field, filling the interior volume, to the Bekenstein-Hawking result. 
The proportionality relations between the entropy of the scalar field in the interior volume of rotating black holes 
and the Bekenstein-Hawking entropy under the Hawking radiation, generalized by considering the particles with energy, 
charge and angular momentum was also investigated recently in \cite{Wang:2019dpk,Wang:2019ear}.

{
In a previous work \cite{Biro:2017flp} we have shown that the proper entropy function of the Schwarzschild black hole 
is actually concave if we treat it as a function of two variables: the mass-energy parameter $E$, and the thermodynamic volume $V$. 
We started our analysis by constituting a power-law equation of state $S \sim E^a V^b$, and contrasted this with the requirement that the temperature (calculated from the thermodynamical derivative of $S$) be equal to the Hawking temperature  \cite{Hawking:1974sw}.
By imposing the physical conditions of 1st order Euler homogeneity and a Stefan-Boltzmann radiation like equation of state at the horizon, we have shown that the thermodynamic volume scales as $ V\sim M^5$.

{In the present work}, we {also} summarize the most important physical concepts behind another thermodynamical volume
definition based on the energy density of Hawking radiation, $\epsilon=\alpha T_{\mathrm{H}}^4$ with the
Hawking temperature $k_{\mathrm{B}}T_{\mathrm{H}}=\frac{\hbar}{2\pi c} \frac{GM}{R^2}$ \cite{Hawking:1974sw} as an average $V=Mc^2/\epsilon$.
Here $\alpha=\gamma\frac{\pi^2}{15} \frac{k_B^4}{(\hbar c)^3}$ is the Stefan--Boltzmann coeffcient
with a degeneracy factor $\gamma$ (e.g.~$2$ for photon radiation).
Based on this sole assumption we derive {again} the scaling law $V \sim M^5$, an extensive entropy with
the related Smarr relation, and a stable equation of state in the next section.


{
In this letter we {also} extend our previous results on the simplest Schwarzschild case to charged and rotating black holes. 
The thermodynamic stability problem of the Kerr-Newman black hole is more complicated than the static 
case because the convexity/concavity of the entropy depends on further thermodynamic parameters.
For example, there is a discontinuity of specific heat at constant angular momentum of the Kerr black hole, where
the sign of the specific heat changes.
Davies claimed \cite{Davies:1978mf} that this discontinuity is an evidence of a certain type of phase transition for rotating black holes.
Later on however, by using the Poincar\'{e} turning point method \cite{poincare1885equilibre}, the stability change of the Kerr black hole was analyzed and it was shown that the argument by Davies is insufficient to ascertain such a phase transition \cite{Kaburaki:1993ah}. A different, zeroth law compatible approach to Kerr black hole thermodynamics applying also the Poincar\'e method was presented by two of us in \cite{Czinner:2017bwc}.
As for the present problem, a part of the Kerr-Newman black hole shows similar thermodynamic behavior to the Schwarzschild case since its entropy 
{ seems to be convex downward.}
Therefore, it is worth to formulate the thermodynamics of the Kerr-Newman black hole 
including the volume term as well. 

The organization of this letter is as follows. 
In section \ref{section:ourapproach}, we explain our previous results on the Schwarzschild black hole {from a different perspective as well}, focusing on the volume scaling and the {thermal} stability 
{aspects of the problem}.
In section \ref{sec:Formulation}, we formulate the volume dependent entropy picture for Kerr-Newman black holes in a similar fashion  as we studied the Schwarzschild {case}, and we derive several thermodynamic variables from the entropy function. In section \ref{sec:Discussion}, we present some discussion of our findings, and summarize the obtained results.
{Throughout this letter} we use units such as $\hbar=G=c=k_{\mathrm{B}}=1$, except in Sec. \ref{section:ourapproach}.
}

\section{{
Volume scaling  and thermodynamic stability}}
\label{section:ourapproach}

{%
Instead of the usual Planck-length and Planck-mass units, {we introduce} two length scales {in this section} associated to the horizon generating total mass, $M$:\\
i.) the Schwarzshild radius, $R(M)=2GM/c^2$, and\\
ii.) the Compton wavelength, $\lambda(M)=\hbar/Mc$.

The former is a purely classical construction {while} the latter includes quantum effects. These will show clearly which quantity is based on semi-classical modifications to classical general relativity.
Note that $R(M) \cdot \lambda(M)=2L_{\mathrm{P}}^2=2\hbar G/c^3$ is independent of $M$, and the
Planck mass is defined by $\lambda(M_{\mathrm{P}})=L_{\mathrm{P}}$.
}

{%
In \cite{Biro:2017flp} we have shown that the entropy of a Schwarzshild black hole
describes a stable (concave) thermodynamics in the 2D parameter space $(E,V)$. Here we repeat
this approach briefly, and in more general terms. For the thermodynamical analysis we need only
two basic facts: {first} that the volume scales like $V \sim M^5$, and 
{second} that the pressure is not zero, and it {does nonzero} work by 
changing this volume. 
Therefore here we start with a short descritpion of the physical characteristica of the Hawking 
radiation, and then list two simple volume concepts related to {this phenomena}.
These and other suggested volumes all follow the very same scaling we choose to fund our conclusions.

The Hawking temperature is an Unruh temperature corresponding to the acceleration called
surface gravity at the classical horizon. We express the corresponding thermal energy here
with help of the two basic length scales:
\be
 k_{\mathrm{B}}T_{\mathrm{H}} \: = \: \frac{\hbar}{2\pi c} \, g \: = \: \frac{1}{4\pi} \, \frac{\lambda(M)}{R(M)} \, Mc^2.
\ee{HAWKTEMPBASIC}
This form shows clearly that in a classical world ($\hbar\to 0$) this would vanish. The
Bekenstein--Hawking entropy is given by
\be
 \frac{1}{k_{\mathrm{B}}}\, S_{\mathrm{BH}} \: = \: \frac{\pi R^2}{L_{\mathrm{P}}^2} \: = \: 2\pi \frac{R(M)}{\lambda(M)}.
\ee{BEKHAWENTROP}
In the classical limit such a quantity tends to infinity. However the product is independent of $\hbar$,
and leads to the Smarr relation
\be
 T_{\mathrm{H}} S_{\mathrm{BH}} \: = \: \frac{1}{2} \, Mc^2.
\ee{HAWSMARR}
Since it does not give the total energy, $E=Mc^2$, the unusual relation, $E=2TS$, became generally
accepted. It is easy to show that adding a mechanical term, as in the whole thermodynamics, can
cure this problem and changes to
\be
 TS \: = \: E \, + \, pV.
\ee{OURSMARR}

First we define the volume as an averaging factor to gain the energy density of a black body radiation
with exactly the Hawking temperature: $\epsilon=Mc^2/V$ leads to
\be
 V \: = \: \frac{Mc^2}{\alpha T_{\mathrm{H}}^4} \: = \: \frac{15}{\gamma} \, (16\pi)^2 \, \frac{R^4}{\lambda}.
\ee{OURVOLUME1}
using $\hbar c = Mc^2 \lambda(M)$. This quantity  clearly has a volume dimension, and analogously to
the Bekenstein--Hawking entropy, its ($\hbar\to 0$) limit is infinite. Substituting the definitions
of the length scales one concludes that $V=\theta \, M^5$. For the thermodynamical consideration
only this scaling and the radiation pressure, $p=\epsilon/3$, is needed.

The entropy is obtained from the thermodynamical integral
\be
  S \: = \: \int \frac{dE+pdV}{T}.
\ee{ENTROPASINTEGRAL}
The first contribution, using the Hawking temperature, is the well known result:
\be
 \int \frac{dE}{T} \: = \: k_{\mathrm{B}} \, \int \frac{c^2dM}{\lambda(M) Mc^2/4\pi R(M)} \: = \: S_{\mathrm{BH}}.
\ee{FIRSTENTROP}
The second contribution is
\be
 \int \frac{pV}{T} \, \frac{dV}{V} \: = \: \int \frac{1}{3} \frac{Mc^2}{T} \, 5 \frac{dM}{M} \: = \: \frac{5}{3} S_{\mathrm{BH}}.
\ee{SECONDENTROP}
The total entropy we consider is therefore $S=8S_{\mathrm{BH}}/3$. This {modifies} the Smarr relation to
\be
 E \: = \: TS - pV \: = \: \frac{8}{3} \, \frac{1}{2} \, Mc^2 \, - \, \frac{1}{3} \, Mc^2 \: = \: Mc^2.
\ee{CORRECTSMARR}
According to this basic relation one in general proves that this entropy is extensive,
it is an Euler-homogeneity class $1$ function: $S(\xi E, \xi V) = \xi S(E,V)$. A consequence,
derivable by taking the differential of this form with respect to $\xi$ and solving the resulted
partial differential equation, is the special functional form $S(E,V)=V s(E/V)$. In our case
\be
 S(E,V) \: = \: \frac{4\sqrt{\pi}}{3} \left(\frac{\gamma}{15} \right)^{1/4} \, k_B (\hbar c)^{-3/4} \:    E^{3/4} V^{1/4}.
\ee{OURMICROEOS}
The above used definition of volume by simulating the total energy content behind the horizon
by a hypothetically observed Hawking radiation may seem arbitrary to some. In order to support
{this} choice {however}, we consider the 
simple {and natural} physical picture, {namely} the evaporation of black 
hole horizons by such a radiation.
Abbreviating the coefficient in the volume defined by us as $V=\nu R^4/\lambda$, the loss of
energy in time is given by
\be
 \frac{d}{dt} Mc^2 \: = \: -4\pi R^2 c \alpha T^4 \: = \: - 4\pi cR^2 \frac{Mc^2}{\nu R^4/\lambda}
\ee{LOSSBHAWKRAD}
Noting that $Mc^2=c^4R/2G$ and $\lambda=2L_{\mathrm{P}}^2/R$ we arrive at the following evolution 
equation for the radius of a Schwarzschild horizon:
\be
 \frac{d}{cdt} \, R \: = \: - \frac{8\pi}{\nu} \, \frac{L_{\mathrm{P}}^2}{R^2}.
\ee{DRDTBH}
From this the evaporation time, $t_{{\rm evap}}=(\nu/8\pi c) \cdot (R^3/3L_P^2)$ follows,
giving rise to an evaporation volume of
\be
 V_{{\rm evap}} \: = \: ct_{{\rm evap}} 4\pi R^2 \: = \: \frac{\nu}{6} \frac{R^5}{L_{\mathrm P}^2} \: = \: \frac{1}{3} V.
\ee{EVAPVOLUME}
This including volume of the evaporation event is in the same order as our thermodynamical volume defined above.
Beyond these intuitive approaches there are {other}, more sophisticated, invariant geometry based volume {definitions} in use, {e.g.~the one by Christodoulou and Rovelli \cite{Christodoulou:2014yia}}. Common in all that $V=\theta M^5$, so the thermodynamical consequences are the same.

Finally we discuss the thermodynamical stability. It is governed by eigenvalues of the second
partial derivative matrix of the microcanonical equation of state, $S(E,V)$. To view this as
a composite $S(M)=S(E(M),V(M))$ function is possible, however the second parametric
derivative against the parameter $M$ in this case also contains contributions with the first
partial derivatives of $S$ and second derivatives of the parametric dependences $E(M)$ and $V(M)$:
\be
 \frac{d^2S}{dM^2} \: = \: \eta(S) \, + \, \pd{S}{E} E^{\prime\prime}(M) \, + \, \pd{S}{V} V^{\prime\prime}(M),
\ee{SECONDDERIV}
with $\eta(S)$ containing the second partial derivatives. These are related to 
transport coefficients, like heat capacity and compressibility. Substituting the thermodynamical
meaning of the first partial derivatives and noting that in our case $E^{\prime\prime}(M)=0$ we arrive
at 
\be
 \eta(S) \: = \: \frac{d^2S}{dM^2} \, - \, \frac{p}{T} \, V^{\prime\prime}(M).
\ee{CORRECTSECOND}
From this formula one immediately realizes that the traditionally considered first term in the stability
description being positive, the second term must overcompensate this in order to stabilize.
Indeed with the Hawking radiation pressure proposed {above as} $p=Mc^2/3V$ and $1/T=1/T_{\mathrm{H}}=2S_{\mathrm{BH}}/Mc^2$,
we arrive at
\be
 \eta(S) \: = \: \frac{16}{3} \, \frac{S_{\mathrm{BH}}}{M^2} \, - \, \frac{40}{3} \, \frac{S_{\mathrm{BH}}}{M^2} \: = \: - 3 \, \frac{S}{M^2} \: < \: 0.
\ee{ETASNOW}
From $S \sim E^{3/4}V^{1/4}$ one draws the same conclusion.
Figure \ref{fig;entropy_Sch} is a plot of the entropy $S$ as a function of $E$ and $V$. 
{ The entropy surface is everywhere convex upward.}
The volume--mass relation $V \sim M^5$ specifies a certain path (blue curve) on this plot. Along this path, the thermodynamics is stable, at any point the surface curvature is {positive.}
However the projection of this path on a plane of constant $V$ (red curve, plotted for $V=0$) becomes 
{convex downward}
and the thermodynamics is unstable on this projection, due to a negative specific heat of the projected entropy.
\begin{figure}
 \begin{center}
  \includegraphics[width=90mm]{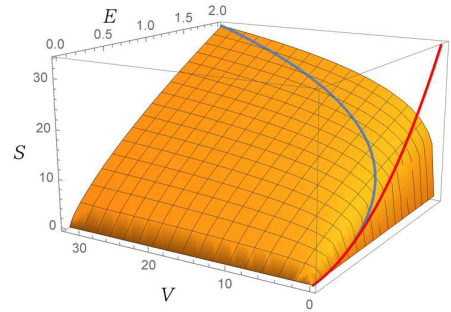}
  \end{center}
  \caption{Plot of the volume dependent entropy $S(E,V)$ for a Schwarzschild back hole. 
	The blue curve is a certain path corresponding to  $V \sim M^5$, while 
	the red curve is its projection onto the $V=0$ plane.}
 \label{fig;entropy_Sch}
\end{figure}

Summarizing this section, with a single natural assumption of considering a volume $V=\epsilon/Mc^2$, associated
to Hawking radiation and correctly taking into account the radiative pressure's work we have derived
the following results: i) such a volume scales like $V \sim M^5$ with the mass parameter, in agreement
with other, independent definitions,
ii) this treatment defines a horizon entropy $S=8S_{\mathrm{BH}}/3$, iii) it leads to a Gibbs potential and
Smarr relation compatible with an extensive $S(E,V) \sim E^{3/4}V^{1/4}$ entropy, and finally
stabilizes the coressponding thermodynamics with $\eta(S)=2S/M^5-5S/M^2 < 0$.
}

\section{Extension to charged and rotating black holes} 
\label{sec:Formulation}
With the angular momentum $J$ and the electric charge $Q$, the entropy of the Kerr-Newman black  hole is given by
\begin{equation}
  S_{\mathrm{BH}} = \sigma ( \mu_1,\mu_2) M^2,
\label{eq:BHentropy}
\end{equation}
with $\mu_1=Q/M$, $\mu_2=J/M^2$ and
\begin{equation}
 \sigma (\mu_1,\mu_2) = 2 \pi \left( 1 + \sqrt{1 - \mu_1^2 - \mu_2^2} - \frac{\mu_1^2}{2} \right).
\end{equation}
It is known that the black hole entropy is a generalized homogeneous function \cite{1972PhRvB...6.3515H} which scales as
\begin{equation}
 \xi^2 S_{\mathrm{BH}}(M, Q, J) = S_{\mathrm{BH}}(\xi M, \xi Q, \xi^2 J).
\label{eq:scale_ent_BH}
\end{equation}
 From (\ref{eq:scale_ent_BH}), we can deduce the Smarr relation \cite{Smarr:1972kt} of the black hole,
\begin{equation}
 M = 2 T_{\mathrm{H}} S_{\mathrm{BH}} + \Phi Q + 2 \Omega J,
\label{eq:Smarr_BH}
\end{equation}
where 
$\Phi$ and $\Omega$ are 
the electric potential at the horizon and the angular velocity of the horizon, 
respectively. 

Here we review the derivation of (\ref{eq:Smarr_BH}) from (\ref{eq:BHentropy}).  
The logarithmic differential of (\ref{eq:BHentropy}) is 
\begin{equation}
  \frac{dS_{\mathrm{BH}}}{S_{\mathrm{BH}}} = 2 \frac{dM}{M} + \frac{\sigma_{,1}}{\sigma} d \mu_1 + \frac{\sigma_{,2}}{\sigma} d \mu_2,
 \label{eq:derivSBH}
\end{equation}
where $\sigma_{,i} = \frac{\partial \sigma}{\partial \mu_i}$.
By using the logarithmic derivatives,
\begin{equation}
 \frac{d\mu_1}{\mu_1} = \frac{dQ}{Q} - \frac{dM}{M}
 \label{eq:dmu1}
\end{equation}
and
\begin{equation}
 \frac{d\mu_2}{\mu_2} = \frac{dJ}{J} - 2\frac{dM}{M},
 \label{eq:dmu2}
\end{equation}
(\ref{eq:derivSBH}) can be rewritten as 
\begin{equation}
  \frac{dS_{\mathrm{BH}}}{S_{\mathrm{BH}}} = \left( 2 - \mu_1\frac{\sigma_{,1}}{\sigma} - 2  \mu_2\frac{\sigma_{,2}}{\sigma} \right) 
	\frac{dM}{M} + \mu_1 \frac{\sigma_{,1}}{\sigma} \frac{dQ}{Q} +  \mu_2 \frac{\sigma_{,2}}{\sigma}
	\frac{dJ}{J} .
\label{eq:derivSBH2}
 \end{equation}
We read off the intensive variables as coefficients in the expression for $dS_{\mathrm{BH}}$ :
\begin{equation}
 \frac{1}{T_{\mathrm{H}}} = \left( 2  - \mu_1\frac{\sigma_{,1}}{\sigma} - 2  \mu_2\frac{\sigma_{,2}}{\sigma} \right) \frac{S_{\mathrm{BH}}}{M},
\end{equation}
\begin{equation}
 - \frac{\Phi}{T_{\mathrm{H}}} = \mu_1  \frac{\sigma_{,1}}{\sigma}  \frac{S_{\mathrm{BH}}}{Q} = \frac{\sigma_{,1}}{\sigma}  \frac{S_{\mathrm{BH}}}{M} 
                                                      = \sigma_{,1} M,
\end{equation}
and
\begin{equation}
 - \frac{\Omega}{T_{\mathrm{H}}} = \mu_2 \frac{\sigma_{,2}}{\sigma}  \frac{S_{\mathrm{BH}}}{J} = \frac{\sigma_{,2}}{\sigma}  \frac{S_{\mathrm{BH}}}{M^2} = \sigma_{,2} .
\end{equation}
One can easily confirm that the Smarr relation (\ref{eq:Smarr_BH}) follows from these equations.

We extend the Bekenstein-Hawking entropy of a Kerr-Newman black hole to the volume dependent version. We assume that this entropy contains general powers of mass $M$ and volume $V$ in the form of,
\begin{equation}
 S = \kappa (\mu_1, \mu_2 )  M^a V^b,
 \label{eq:volS}
\end{equation}
{%
as an extension of Schwarzschild case.
}
Obviously, $\kappa(\mu_1,\mu_2)$ must differ from $\sigma(\mu_1,\mu_2)$.
The scaling relation of this entropy function,  
\begin{equation}
  \xi^{a+b} S(M, V, Q, J) = S(\xi M, \xi V, \xi Q, \xi^2 J),
\end{equation}
generates the following Smarr-like relation:
\begin{equation}
 M = (a + b) T S - p V + \Phi Q + 2 \Omega J.
\label{eq:Smarr}
\end{equation}
The logarithmic differential of (\ref{eq:volS}) is now given by
\begin{equation}
  \frac{dS}{S} = a \frac{dM}{M} + b \frac{dV}{V} + \frac{\kappa_{,1}}{\kappa} d\mu_1 + \frac{\kappa_{,2}}{\kappa}d\mu_2.
\end{equation}
Inserting $d \mu_1$ and $d \mu_2$ from (\ref{eq:dmu1}) and (\ref{eq:dmu2}), we obtain
\begin{eqnarray}
	\frac{dS}{S} &=& \left( a - \mu_1 \frac{\kappa_{,1}}{\kappa} - 2 \mu_2 \frac{\kappa_{,2}}{\kappa}  \right) \frac{dM}{M}\nonumber\\ 
	&+& b \frac{dV}{V}  + \mu_1 \frac{\kappa_{,1}}{\kappa} \frac{dQ}{Q} + \mu_2 \frac{\kappa_{,2}}{\kappa} \frac{dJ}{J}.
\end{eqnarray}
One easily extracts the temperature $T$, the pressure $p$, the electromagnetic potential $\Phi$, and the angular velocity $\Omega$, as follows
\begin{eqnarray}
  \frac{1}{T} &=& \left( a - \mu_1 \frac{\kappa_{,1}}{\kappa} - 2 \mu_2 \frac{\kappa_{,2}}{\kappa}  \right)  \frac{S}{M},\\
 \frac{p}{T} &=& b\ \frac{S}{V},\\
 - \frac{\Phi}{T} &=& \mu_1 \frac{\kappa_{,1}}{\kappa} \frac{S}{Q} = \frac{\kappa_{,1}}{\kappa} \frac{S}{M},\\
 - \frac{\Omega}{T} &=& \mu_2 \frac{\kappa_{,2}}{\kappa} \frac{S}{J} = \frac{\kappa_{,2}}{\kappa} \frac{S}{M^2}.
\end{eqnarray}
One can verify that the extended Smarr relation (\ref{eq:Smarr}) is deduced from the above relations.
{
Also, one can show that these variables satisfy the first law with work term,
\begin{equation}
 TdS = dM + p dV -\Phi dQ - \Omega dJ.
\end{equation}
}

In what follows we derive expressions for the volume $V$ and the pressure $p$ in the thermodynamic system constructed by the above entropy function $S(M,V,Q,J)$.
First we assume that the temperature $T$ is equal to $T_{\mathrm{H}}$: 
\begin{equation}
  \left( 2  - \mu_1\frac{\sigma_{,1}}{\sigma} - 2  \mu_2\frac{\sigma_{,2}}{\sigma} \right) \frac{S_{\mathrm{BH}}}{M} = \left( a - \mu_1 \frac{\kappa_{,1}}{\kappa} - 2 \mu_2 \frac{\kappa_{,2}}{\kappa}  \right)  \frac{S}{M}.
\end{equation}
Next we equate the other intensive variables $\Phi$ and $\Omega$ of both systems and obtain the following relations,
\begin{equation}
 \frac{\sigma_{,1}}{\sigma}  \frac{S_{\mathrm{BH}}}{M}  = \frac{\kappa_{,1}}{\kappa} \frac{S}{M} ,
\label{eq:s1k1}
\end{equation}
and
\begin{equation}
 \frac{\sigma_{,2}}{\sigma}  \frac{S_{\mathrm{BH}}}{M^2} = \frac{\kappa_{,2}}{\kappa} \frac{S}{M^2}.
\label{eq:s2k2}
\end{equation}
A simple consequence of these three equations is the proportionality of the entropy functions,
\begin{equation}
	2 \, S_{\mathrm{BH}} \: = \: a \, S.  
\end{equation}
Substituting this result into (\ref{eq:s1k1}) and (\ref{eq:s2k2}), we obtain the differential equation
\begin{equation}
 \frac{\kappa_{,i}}{\kappa} = \frac{a}{2} \frac{\sigma_{,i}}{\sigma},
\end{equation}
where $i = 1, 2$. The solution to this equation is 
\begin{equation}
 \kappa = \kappa_0 \sigma^{\frac{a}{2}},
\end{equation}
where $\kappa_0$ is a constant.
This result leads us to the following expression of the volume,
\begin{equation}
 \label{eq:volumeKN}
 V = \left( \frac{S}{\kappa M^a} \right)^{\frac{1}{b}} = \left( \frac{1}{\kappa_0} \frac{2}{a} \right)^{\frac{1}{b}} \left(  \sigma M^2 \right)^{\frac{1}{b}\left( 1 - \frac{a}{2} \right)}.
\end{equation}
As a result, we conclude that the thermodynamic volume is proportional to $M^{\frac{2-a}{b}}$.
The corresponding expression for the pressure is then obtained as,
\begin{equation}
 p = b \frac{ST}{V} = \frac{b}{a} \frac{M}{V} \sqrt{1 - \mu_1^2 - \mu_2^2},
\end{equation}
where we use
\begin{equation}
  T = T_{\mathrm{H}} = \frac{\sqrt{1 - \mu_1^2 - \mu_2^2}}{2 \sigma M}.
\end{equation}
It is clear that both the pressure and the temperature vanish in the limit of
an extremal Kerr-Newman black hole.
{
Here we equated the temperature with the Hawking temperature, which is the one of thermal radiation called Hawking radiation.
Therefore one can treat the pressure given in the above expression as a thermal radiation pressure which would be of quantum mechanical origin.
}

In \cite{Biro:2017flp} we have shown that, in the Schwarzschild case, only one possible solution, 
$a=\frac{3}{4}$ and $b=\frac{1}{4}$ emerges assuming  Euler-homogeneity, $a+b=1$, 
of the entropy and a Stefan-Boltzmann radiation like equation of state in 3 spatial dimensions, $b=a/3$.  
In the present Kerr-Newman case, the Schwarzschild limit is achieved for $\mu_1 = \mu_2 =0$. 
So it is natural to demand that the volume dependent entropy of the Kerr-Newman black hole
has the same scaling powers as the Schwarzschild one. As a result, the scaling of the entropy 
occurs as
\begin{equation}
 \label{eq:lambdaS}
 \xi S(M, V, Q, J) = S(\xi M, \xi V, \xi Q, \xi^2 J),
\end{equation}
and the proportionality relation between the entropy functions becomes,
\begin{equation}
 S = \frac{8}{3} S_{\mathrm{BH}}.
\end{equation}
The thermodynamic volume is proportional to $M^5$, just as in the Schwarzschild limit. 
The general expression for the Kerr-Newman horizon pressure 
has then its maximum at one third of the energy density $M/V$:
\begin{equation}
p =\frac{1}{3}  \frac{M}{V} \sqrt{1 - \mu_1^2 - \mu_2^2}.
\end{equation}


\section{Discussion and summary}
\label{sec:Discussion}
We have shown that the thermodynamic volume $V$, of a Kerr-Newman black hole is proportional to the power of mass energy $M$ with the 
power index $ (2-a)/b$. 
The scaling becomes $V \sim M^5$ when the static black hole limit corresponds to the one derived from the 
assumptions of the 1st order Euler homogeneity and the Stefan-Boltzmann radiation like equation of state at the horizon. 
It is very interesting to mention that by requiring that the loss
in the internal mass energy and the mechanical work show the same scaling, $dM \sim pdV$, the Hawking radiation
pressure, $p\sim T^4 \sim M^{-4}$, leads directly to $dM \sim M^{-4}dV$, from which $V(M)\sim M^5$
trivially follows. On the other hand inspecting the energy loss by the Hawking radiation through a surface of $A \sim M^2$
one obtains \hbox{$dM \sim A T^4 dt \sim M^{-2} dt$} and due to this $dV \sim A dt \sim M^4 dM$, which leads to the same result.
For the Schwarzschild black hole, this volume scaling agrees with the estimation from the CR-volume considering 
the black hole evaporation by Hawking radiation.
For the axially symmetric black hole, it is difficult to calculate the interior volume based on the method proposed 
by Christodoulou and Rovelli because the largest spacelike hypersurface, filling the region behind the horizon,
cannot be found easily.
Some authors have given only approximate evaluations of Kerr and Kerr-Newman black holes \cite{Bengtsson:2015zda,Wang:2019ake}.
Also, while there are qualitative investigations about the Hawking radiation with angular momentum 
\cite{Wang:2019dpk,Wang:2019ear},
there is an ambiguity in the evaporation process of rotating black holes 
because we do not know how much angular momentum can be emitted. This also must depend on the
polarization of the radiation.
Therefore, it is difficult to say anything about the relation between the volume scaling 
obtained here and the CR-volume of rotating and possibly charged black holes.

In summary, in this letter we have extended the volume dependent entropy of a Schwarzschild black hole to an axially symmetric rotating and charged black hole.
With the assumption that the entropy depends on the power of volume, we have derived the expressions for the volume and the pressure in the thermodynamic system constructed.
It has been demonstrated and supported by physical arguments, that the thermodynamic 
volume is proportional to the fifth power of the mass-energy parameter, and
the derived pressure is proportional to the energy density, i.e.~$M/V$.

We have pointed out that the thermodynamic stability of such horizons requires a sublinear
volume dependence whenever Euler-homogeneity (thermodynamical extensivity) is cared for.
The special value $b=1/4$ emerges from the $3D$ isotropic radiation pressure,
although it is a valid question how accurate this assumption can be in an axially symmetric 
system. With this assumption however, the scaling of the volume becomes $V \sim M^5$, just 
like in the static case, and agrees with the CR-volume scaling. While the CR-volume scaling has 
not yet been known for the axially symmetric case, we expect that the extensivity of the entropy 
function is governed solely by the volume and the total mass-energy dependence.
As long as $a+b=1$ is fulfilled, the rest depends only on the specific ratios, 
like $J/M^2$ and $Q/M$. The stability depends on the same factors as it does in the Schwarzschild 
case.

As for the Euler-homogeneity of the entropy, the extended entropy seems to be still a generalized homogeneous function, just like the nonextended one. Even so, it may be possible to 
argue that the volume dependent entropy is a 1st order Euler homogeneous function if we consider the 
finite angular momentum case. If so, even the general black hole is thermally stable within this approach.

\section*{Acknowledgements}
This work was supported by the Hungarian National Bureau for Development, Innovation and Research (NKFIH) under
the project number K123815.


\bibliography{Kerr_ref}
\bibliographystyle{elsarticle-num}

\end{document}